\documentclass[pre,twocolumn,showpacs,amsmath,superscriptaddress,amssymb]{revtex4}
\usepackage[dvips]{graphicx}
\usepackage{dcolumn}

\begin{document}
\title{Dynamics of Nano-Confined Water under Pressure}
\author{S.O. Diallo}
\email{omardiallos@ornl.gov}
\affiliation{Quantum Condensed Matter Division,  Oak Ridge National Laboratory, Oak Ridge, Tennessee 37831, USA}
\author{M. Ja\.zd\.zewska}
\affiliation{Faculty of Physics, Adam Mickiewicz University, 61-614 Pozn\'an, Poland}
\author{J.C. Palmer}
\affiliation{Department of Chemical and Biomolecular Engineering, North Carolina State University, Raleigh, North Carolina 27695, USA}
\author{E. Mamontov}
\affiliation{Chemical and Engineering Science Division,  Oak Ridge National Laboratory, Oak Ridge, Tennessee 37831, USA}
\author{K.E. Gubbins}
\affiliation{Department of Chemical and Biomolecular Engineering, North Carolina State University, Raleigh, North Carolina 27695, USA}
\author{M. \'Sliwi\'nska-Bartkowiak}
\email{msb@amu.edu.pl}
\affiliation{Faculty of Physics, Adam Mickiewicz University, 61-614 Pozn\'an, Poland}

\begin{abstract}
We report a study of the effects of pressure on the diffusivity of water molecules confined in single-wall carbon nanotubes (SWNT) with average mean pore diameter of $\sim$ 16 {\AA}.  The measurements were carried out using high-resolution neutron scattering, over the temperature range 220 $\le T\le$ 260 K, and at two pressure conditions: ambient and elevated pressure. The high pressure data were collected at constant volume on cooling, with $P$ varying from $\sim$1.92 kbar at temperature $T=$ 260 K  to $\sim$1.85 kbar at $T=$ 220 K. Analysis of the observed dynamic structure factor $S(Q,E)$ reveals the presence of two relaxation processes, a faster diffusion component (FC) associated with the motion of \lq caged' or restricted molecules, and a slower component arising  from the free water molecules diffusing within the SWNT matrix. While the temperature dependence of the slow relaxation time exhibits a Vogel-Fulcher-Tammann law and is non-Arrhenius in nature, the faster component follows an Arrhenius exponential law at both pressure conditions. The application of pressure remarkably slows down the overall molecular dynamics, in agreement with previous observations, but most notably affects  the slow relaxation. The faster relaxation shows marginal or no change with pressure within the experimental conditions.

\end{abstract}
\maketitle

\date{\today}

\section{Introduction}
Fluids confined in tight cavities, or between mineral interfaces or biological molecules are ubiquitous in nature. In particular,  water confined in restricted spaces appears to be present in many relevant life situations on the surface of the earth and in our bodies \cite{Stanley_book,Zaccai:00}. This confined water exhibits unusual physical properties, that are generally different than the bulk. Understanding these properties and their connections with life is therefore of great fundamental value \cite{Stanley_book,Chaplin:09}. 

Despite much efforts \cite{Takaiwa:08,Kyakuno:11}, our knowledge about confined water  at ambient conditions - let alone under pressure - remains rather limited. To date, there is still no well established global pressure-temperature $(P,T)$ phase diagram, for varying confine sizes. In contrast,  bulk water has a rather well known $P$-$T$ diagram, spanning a wide range of $P$ and $T$ points. This diagram reveals that many crystalline forms of water are only found at pressures well above 1 kbar, and/or at temperatures below 200 K, and therefore are not easily accessible under normal experimental conditions.   Thanks to recent advances in synthesis of novel nanomaterials such as carbon nanotubes (SWNT) \cite{Iijima:91,Kresge:92},  there are indications that these ice phases can also occur near ambient conditions under confinement \cite{Takaiwa:08,Matsuda:06}.   
\begin{figure}
\includegraphics[width=3.2in,angle=-0]{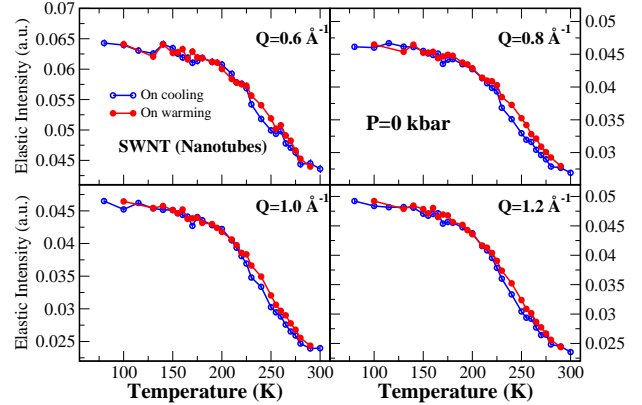}
\caption{(Color) Temperature and wavevector dependence of the elastically scattered neutron intensity obtained from the hydrated SWNT sample,  as measured on cooling and heating. Crystallization is typically characterized by an abrupt change in the elastic intensity, which is not observed at any wavevector transfer $Q$. Confining water in very small pores clearly suppresses crystallization to much lower temperatures, and allows water molecules to remain mobile down to very low temperatures. }
\label{fig1}
\end{figure}
While there has been a considerable amount of work on  the dynamics of interfacial and confined water at ambient pressure \cite{Zanotti:99a,Zanotti:05,Mamontov:06,Chu:07},  few comparable studies have been conducted  at high pressure, due largely to experimental hurdles. With the exception of a few recent reports \cite{Liu:06,Matsuda:06,Kyakuno:11}, the dynamics of confined water as a function of pressure remains largely unexplored.  In this work, we report a quasielastic neutron scattering (QENS) study aimed at investigating the effects of external pressure on the diffusion and molecular dynamics of water adsorbed on commercially available hydrophobic single-wall carbon nanotubes (SWNT) in the temperature range between 220 and 275 K. The observed QENS spectra reveal the existence of two different relaxation processes, which are clearly separated in time by 1-2 orders of magnitude. The broadening in energy of both processes can be described by a liquid like jump diffusion model. The  relaxation times of the faster process exhibit an Arrhenius temperature dependence, while those of the slow component follows a Vogel-Fulcher-Tammann law. We find that the application of pressure slows down the overall molecular dynamics of water in SWNT, consistent with previous studies in hydrophilic porous silica \cite{Liu:05,Liu:06}. Up to the maximum working pressure of our experiment ($P\leq2$ kbar),we find that the faster component is largely unaffected by pressure, and the slow diffusing component is significantly influenced by pressure. 

\begin{figure}
\includegraphics[width=3.2in,angle=0]{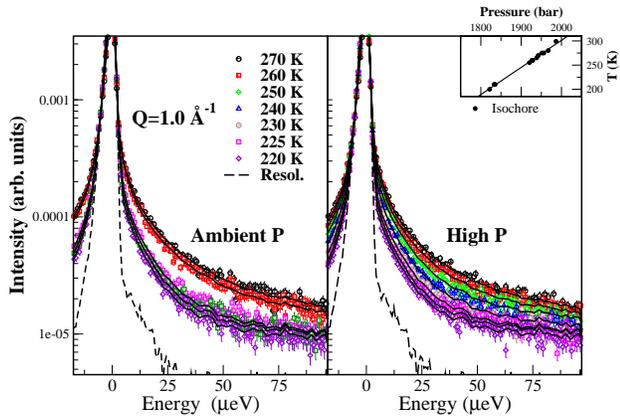}
\caption{(Color) Temperature dependence of the quasielastic response of  water confined in16 {\AA} SWNT. The solid symbols are the experimental data and the solid black lines are model fits, as described in the text. The instrument resolution function measured with the same exact sample at 30 K is shown for comparison (dashed line).  Ambient pressure data are shown on the left hand side, while the high pressure data are shown on the right hand side. The inset shows the isochore along which the measurements were taken.}
\label{fig2}
\end{figure}

\section{Experimental details}
As introduced above, we used a commercially available SWNT sample (purchased from Nanocyl in  Belgium) to investigate the diffusion of water under pressure. Carbon nanotubes are molecular channel of graphitic carbon with remarkable properties and vast potential future applications, including hydrogen storage and molecular separation.  SWNT form simple nano-channels that are on average very similar in both size and hydrophobic character to biological channels, and can be filled with water at ambient conditions.  Our open-ended SWNT sample has an average pore diameter of 16 \AA. The present sample, which was produced via the catalytic carbon vapor deposition process,  has been characterized by the manufacturer using various  techniques such as small angle x-ray scattering (SAXS), high resolution microscopy (TEM), Raman spectroscopy and nitrogen adsorption isotherms. The estimated surface area is  slightly over $~$1000 $m^2$/g.

\begin{figure}
\includegraphics[width=3.2in]{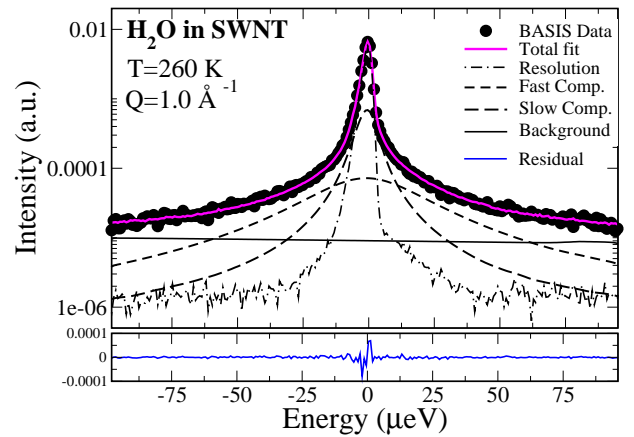}
\caption{(Color) Data and fitting model. QENS signal observed at $T=$260 K and at wavevector $Q=1$ \AA$^{-1}$ for water in 16 {\AA} SWNT (black solid circles). The solid magenta line is the overall fitting function, described in the text. The broad Lorentzian component, with characteristic HWHM as $\Gamma_1(Q)$, is depicted by the short dashed line while the narrow Lorentzian function, with characteristic HWHM $\Gamma_2(Q)$, by the long dashed line. The solid black line represents the sloping background, and the dash-dotted line is the instrumental resolution function. To highlight the goodness of the fit, the difference between the data and the fit, denoted residual (blue solid line), is shown at the bottom.}
\label{fig3}
\end{figure}

\subsection{Sample Preparation} 
Prior to the measurements, the sample was dried for 48 hours under vacuum at 358 K. The dried sample was then hydrated in a humid atmosphere at room temperature for several hours until its mass increases by about 10\%.  The hydrated sample was then transferred into a specially designed cylindrical aluminum cell for high pressure measurements.  While the cell used was nominally rated to 5 kbar for safe operation, all our measurements were done below 2 kbar to comply with safety due to use cycle of the cell. A new cell is currently being developed for future experiments at high pressures.

\begin{figure*}
\includegraphics[width=5in,angle=0]{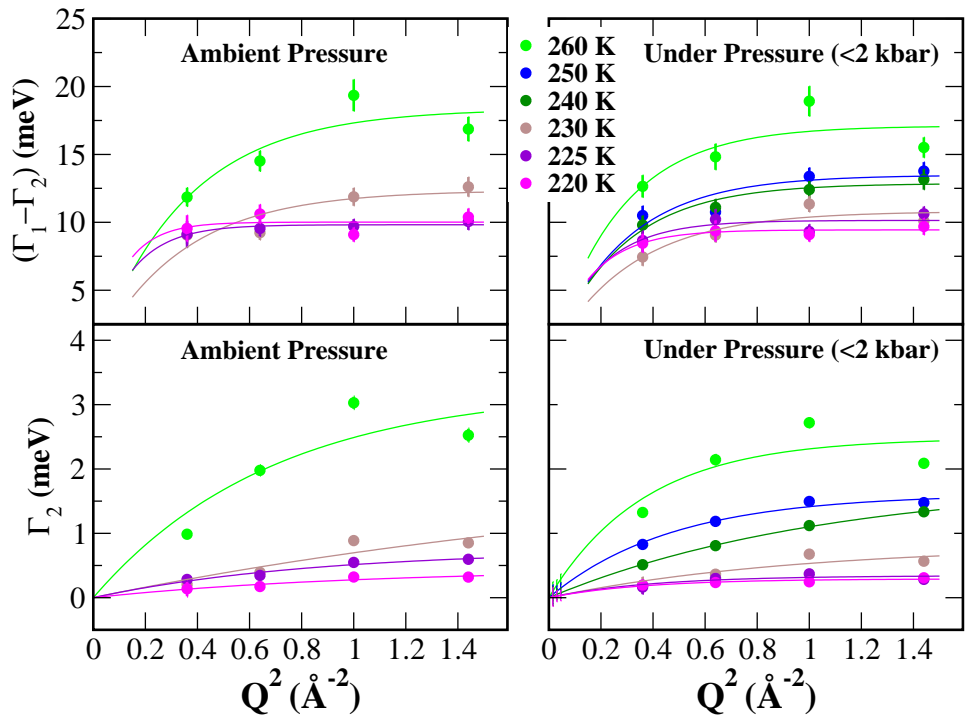}
\caption{(Color) Temperature evolution of the observed Lorentzian HWHMs as a function of $Q^2$.  The faster diffusion component is characterized by the parameter $\Gamma_1(Q)-\Gamma_2(Q)$ and the slower component by $\Gamma_2(Q)$. (LHS): Ambient pressure data. (RHS): High pressure data. The difference parameter ($\Gamma_1-\Gamma_2$)  accounts for the small coupling between the two diffusion processes in the time domain, as discussed for example in \cite{Mamontov:09,Qvist:11}. }
\label{fig4}
\end{figure*}

\subsection{Neutron Experiment}
The high resolution neutron scattering experiments were carried out using the Backscattering Silicon Spectrometer (BASIS) at the 1.4 MW Spallation Neutron Source (SNS), Oak Ridge National Laboratory (ORNL) \cite{Mamontov:11}. BASIS was selected for the present study because of its unique wide dynamic range $\Delta$E=$\pm$100 $\mu$eV and its excellent energy resolution of 1.75 $\mu$eV (Half-Width at Half Maximum or HWHM) at the elastic position. 
We begin our measurements with standard short  \lq elastic intensity' scans on the hydrated SWNT sample at ambient pressure. The goal was to determine a suitable temperature region  for subsequent long QENS measurements, which generally requires high statistics. The elastically scattered neutrons  were recorded over a wider temperature range fairly quickly, as they do not require high count rates. Data were collected with relatively small temperature increments on cooling from 300 K to 80 K. Fig. \ref{fig1} show the raw elastic intensity as a function of temperature for the $Q$ investigated here. The elastic intensity for each temperature was obtained by integrating the corresponding spectrum over a very narrow energy range of ${\pm}$3.5 $\mu$eV, corresponding to the elastic resolution. For an isotropic system, we anticipate the elastic intensity to have a Debye-Waller behavior, i.e. $\propto \exp{(-Q^2\langle u^2(T)\rangle/3)}$, where $\langle u^2(T)\rangle$ is the mean square amplitude vibration. As the sample cools down,  the molecular diffusion also start to slow down and $\langle u^2(T)\rangle$ decreases. The elastic intensity  within  the 3.5 $\mu$eV energy resolution  effectively increases with decreasing temperature until it reaches a maximum (plateau region at low temperatures). Crystallization is typically characterized by an abrupt change in the elastic intensity, which is not observed here at any wavevector $Q$. This suggests that there is no bulk water present and that the water molecules inside the pores remain mobile down to very low temperatures.

The QENS spectra were recorded in the wavevector transfer range $Q$, 0.6$\leq Q\leq$1.2  {\AA}$^{-1}$, in step $\Delta Q=$ 0.2 {\AA$^{-1}$},  spanning a temperature range from $T=$ 220  to  260 K, and two pressure conditions: atmospheric and elevated. QENS Measurements were first taken under  ambient conditions (with a total of 5 temperature points), followed by the measurements at high pressure (7 points in total). Due to the limited allocated time on the instrument, we were forced to reduce the measurement time at the ambient pressure condition for which previous measurements using the HFBS instrument at NIST have been reported \cite{Mamontov:06,Chu:07}. The high pressure measurements were performed at constant volume, starting from $\simeq$ 1.92 kbar at 260 K  to about 1.85 kbar at 220 K. To achieve the desired pressure, we use a helium gas panel with an intensifier to increase the pressure inside a specially designed Al cell to approximately 2 kbar at 300 K, and seal the cell for the rest of the experiment. Data was then collected on cooling along the isochore, as indicated in the inset of Fig. {\ref{fig2}}.  A major depression of the freezing point for nano-confined water in carbons is expected, due to the hydrophobic nature of the water-carbon interaction.  For such systems the microscopic wetting parameter, $\alpha$, which measures the ratio of the water-carbon to the water-water interaction, is only about 0.5\cite{Alba:06}.

\begin{figure*}
\begin{tabular}{c c}
\includegraphics[width=3.0in]{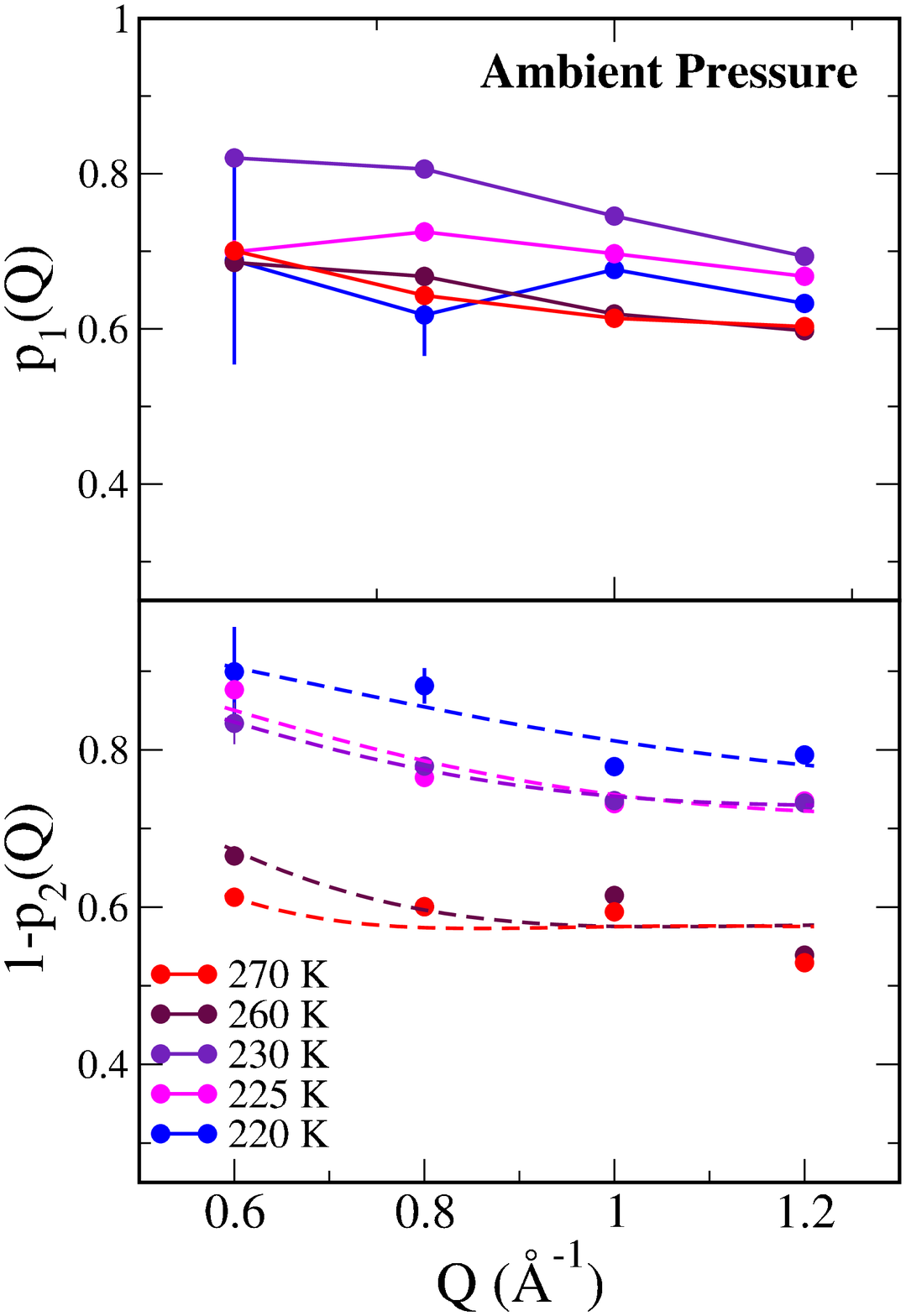}&
\includegraphics[width=3.0in]{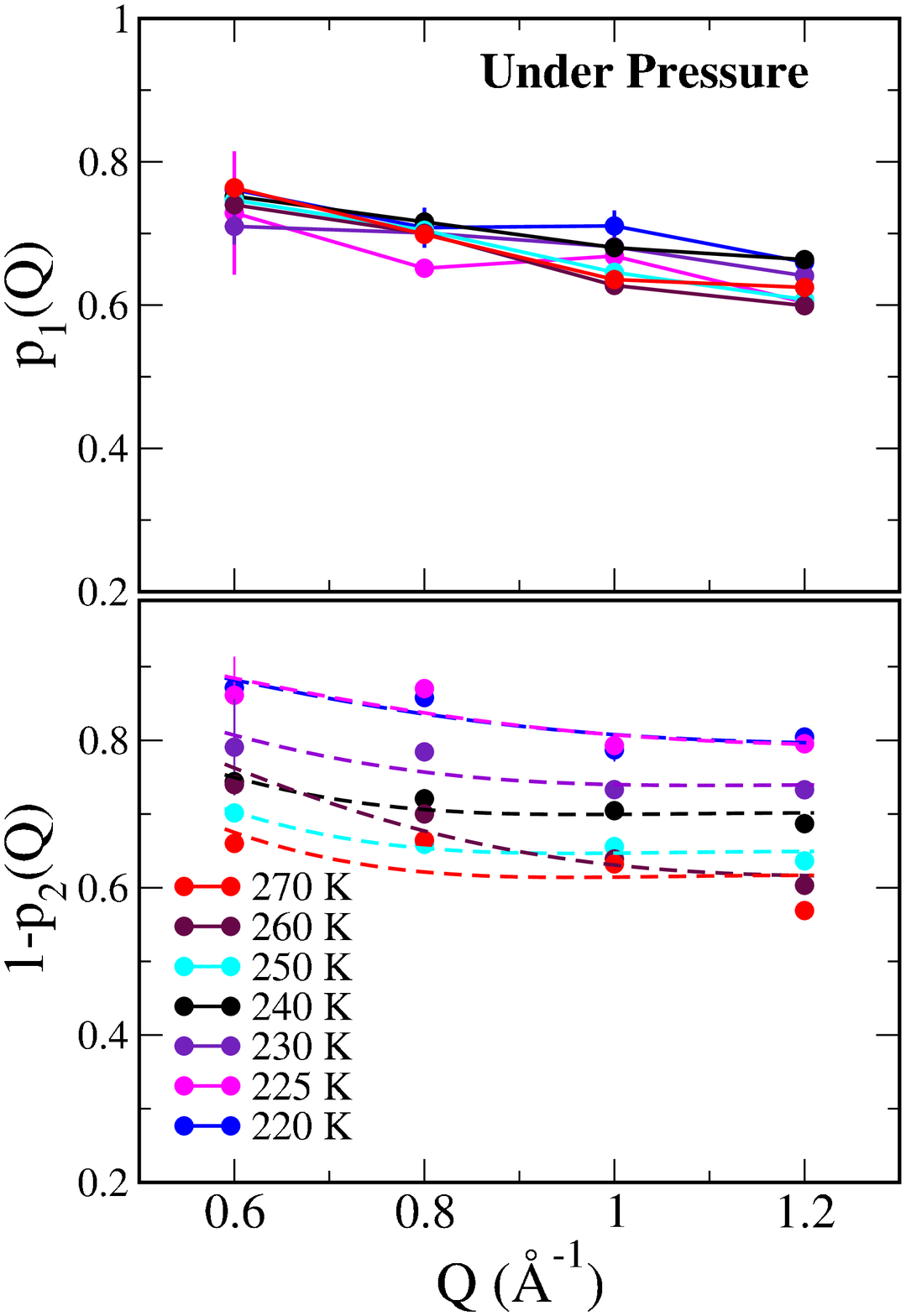}
\end{tabular}
\caption{(Color online) Temperature dependence of the parameters $p1(Q)$ and $1-p_2(Q)$ at ambient and elevated pressure (left and right panel, respectively). The solid symbols are the experimental data and the solid lines are a guide to the eye. The dashed lines are representative model fits based on the restricted transient confinement, as discussed in the text.}
\label{fig5}
\end{figure*}

\section{Results and Discussion} 
The observed temperature dependence of the QENS spectra collected at BASIS  at the conditions  investigated (ambient and high pressure) for wavevector transfer $Q=1.0$ {\AA}$^{-1}$, are  shown side-by-side in Fig. \ref{fig2}.  The filled circles are the experimental data and the overlaid solid lines are model fits, following the fitting procedure which we describe below. The dashed line is the instrument resolution measured using the exact same sample at $T=30$ K where all molecular motions in the sample are expected to become immobile. As anticipated, the QENS broadening narrows as the temperature is reduced, indicating a slowing down of the molecular motion. The neutron scattering spectra $I(Q,E)$ were analyzed using a model scattering function $S(Q,E)$, plus an elastic term $p_1(Q)$ due to all immobile atoms, and a linear background term $B(Q,E)=a+bE$, 
 \begin{equation}
  I(Q,E)=A(Q)[ p_1(Q)\delta(E)+(1-p_1(Q))S(Q,E)+B(Q,E)]
 \label{eq0}
 \end{equation}
 \noindent convoluted with the measured instrument resolution. Our model $S(Q,E)$ include two Lorentzians,
 
 \begin{equation}
S(Q,E)= p_2(Q)\frac{1}{\pi}\frac{\Gamma_1(Q)}{\Gamma_1^2(Q)+E^2}+(1-p_2(Q))\frac{1}{\pi}\frac{\Gamma_2(Q)}{\Gamma_2(Q)^2+E^2} 
\label{eq1}
\end{equation}

We attribute the broader of the two Lorentzians in Eq. \ref{eq1} to the \lq caged' or restricted motion of water molecules, with $\Gamma_1(Q)$ as a main characteristic HWHM. Similarly, we associate the narrow component to the \lq cage-breaking' water molecules with  $\Gamma_2(Q)$ as a characteristic HWHM, and $1-p_2(Q)$ as its relative weight in Eq. \ref{eq1}. Recent compelling arguments for using this two-Lorentzians model over the more traditionally used stretched exponential model has been put forward by Qvist {\it et al.} \cite{Qvist:11}. In their work, Qvist {\it et al.} argued that the faster component in water is most appropriately described as Ô\lq intra-basin' dynamics of the center of mass, a spatially restricted or \lq caged' diffusion which is not strictly rotational. The slower translational component is associated with the \lq inter-basin diffusion jumps, as water molecules in a \lq basin' perform a number of \lq intra-basin' jumps ($\beta$-fast relaxation) before eventually moving ($\alpha$-relaxation, or translational diffusion) to become associated with another \lq basin'.  Eqs. \ref{eq0} and \ref{eq1} yield excellent fits to the data, as depicted in Fig.\ref{fig2}, and as illustrated in Fig.\ref{fig3} for selected temperature and $Q$ value.  The variation of the observed broadenings with $Q^2$ and temperature, is summarized in Fig. \ref{fig4} for both experimental pressure conditions. The behavior of $\Gamma$ with $Q^2$ suggests a jump diffusion process with a distribution of jump length. We thus fit the observed $\Gamma(Q)$ at each $(P,T)$ point with the following expression, \cite{Hall:81}
\begin{equation}
\Gamma(Q)=\frac{\hbar}{\tau}\large[1-\exp(-DQ^2\tau)\large]
\label{eq2}
\end{equation}

\noindent The parameter $\tau$ is the average residence time between jumps, and $D$ the diffusion coefficient.  These two parameters are inversely related via $D=\langle r^2\rangle/6\tau$, where $\langle r^2\rangle$ is the mean squared diffusion jump length. The lines in Fig. \ref{fig4} represent the best fits to the data  using Eq.\ref{eq2}. The diffusion coefficient $D$ is generally best determined at low $Q$, while the residence time $\tau$ is provided by the high $Q$ limit of Eq. \ref{eq2}. Because of this, and given the limited accessible low $Q$ values in our experiment, we focus solely on the influence of pressure on the observed residence time $\tau$. The reported residence time for the faster component has been corrected for the coupling between the two diffusion processes in the time domain (as outlined in { \cite{Mamontov:09}).

\begin{table}
\caption{Characteristic fit parameters for the two observed diffusion components, and their evolution with pressure.}
\begin{ruledtabular}
\begin{tabular}{c |c|c|c|c||c|c|}
Comp. &  Pressure                              & $\tau_0$(ps)   & $T_0$(K) & $F$    & $\tau_A$(ps) &  $E_A$(kJ/mol)\\
\hline
\hline
SLOW   &Ambient                     &  125& 212           &0.09     & -             &-\\
               & &&&&& \\
               & High                           & 80 & 199            & 0.35 & -                   & -  \\
\hline
\hline
FAST   &Ambient                     & - &  -        &-  & 1.04 &7.63\\
               & &&&&& \\
               & High                           &  -& -            & - & 2.01               & 6.49    \\
\end{tabular}
\end{ruledtabular}
\label{tbl1}
\end{table}

The temperature dependence of the observed spectral weight of the overall elastic scattering $p_1(Q)$ and the relative fraction of the narrow Lorentzian $1-p_2(Q)$, introduced in in \ref{eq1},  is summarized in \ref{fig5}. To estimate the radius $a$ of the confining transient cage, we fitted $1-p_2(Q)$ using the expression $f+(1-f)\big(3j_1(Qa)/Qa\big)^2$ \cite{Qvist:11}, where $j_1(x)$ is the spherical Bessel function of the first order, and $f$ the \lq immobile' fraction. From these fits (shown as dashed lines in \ref{fig5}), we find  $a$ to vary from $\sim$ 5.3 \AA at 270 K and ambient pressure (4.7 with pressure) to $\sim$ 2.58 {\AA}  (3.3 under pressure) at 220 K. The errorbars  are somewhat large at low temperatures. 

Fig. \ref{fig6} shows the variation of the residence time $\tau$ with temperature. The faster relaxation exhibits an Arrhenius temperature dependence of the form $\tau=\tau_{A}e^{\frac{E_{A}}{RT}}$,  where the parameters $\tau_A$, and $E_A$ are respectively, a pre-factor and the process activation energy. The slow relaxation follows a Vogel-Fulcher-Tammann (VFT) law $\tau=\tau_0e^\frac{FT_0}{T-T_0}$, where $\tau_0$, $F$, and $T_0$ are respectively, a pre-factor, the fragility parameter, and the ideal glass transition temperature. The resulting fit parameters are summarized in Table \ref{tbl1}. From inspection of Fig. \ref{fig6}, it is evident that within the experimental pressure range probed ($ \le2$ kbar), the faster component is only marginally affected by pressure while changes in slow component are noticeably large. 

The activation energy  $E_A$ for the fast component of nanoconfined water is largely unaffected by pressure, with $E_A=$ 6-7 kJ/mol, a value  somewhat lower than the one obtained for the bulk liquid at comparable temperatures. This lower value is likely due to the interplay between the confining matrix and the hydrogen bonds network. The limited number of hydrogen bonding in confinement, coupled with the interaction of the water molecules with the substrate lattice,  tends to facilitate the diffusion process of water molecules in confinement.

 Interestingly, only the observed VFT parameters associated with the slow component vary significantly with pressure. The fragility parameter $F$ changes by nearly a factor of 4  while the pre-factor $\tau_0$ decreases by approximately 50\% when pressure is applied. Yet, within the precision and the temperature range of our measurements, the observed $\tau$ are consistently larger under pressure. Therefore, additional data points at higher temperatures would be required before an accurate $\tau_0$ can be reported. On the other hand, the parameter $T_0$ is reliably determined, and not affected much by pressure. The observed $T_0$ values are in excellent agreement with previous observations of water in the nanoporous silica family of comparable pore diameter \cite{Liu:05}. It is worth noting that the effect of pressure due to the curved meniscus is negligible and marginally smaller (orders of magnitude) than that of the externally applied pressure \cite{Webber:10}.

\begin{figure}
\includegraphics[width=3.2in]{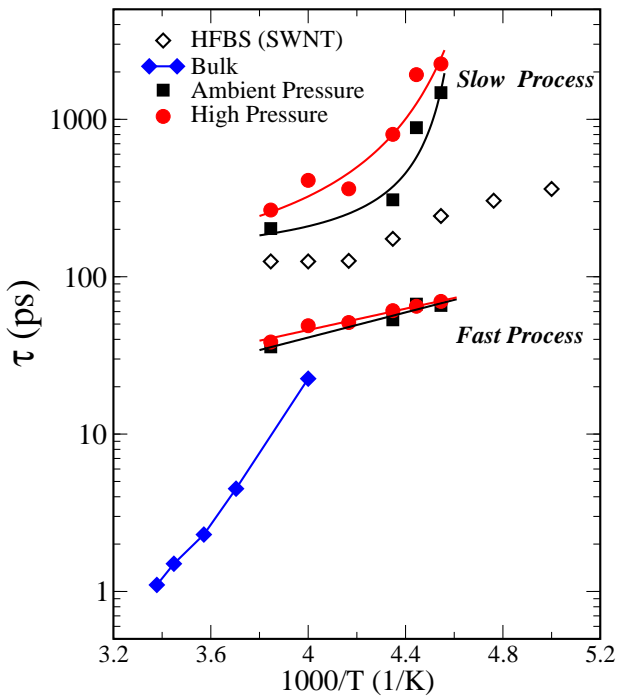}
\caption{(Color) Temperature dependence of the residence time of water confined in 16 {\AA} SWNT (solid squares and circles). The application of external pressure (up to 2 kbar) decreases the overall mobility of water molecules, with a notably stronger effect on the slower diffusion component. The fast component is marginally affected by pressure. Molecular diffusivity in bulk water is faster than in SWNT \cite{Teixeira:85}. The observed $\tau$  values for water confined in 14 {\AA} SWNT, as measured on the high flux backscattering instrument (HFBS) at NIST, are also shown for comparison  (from Ref. \cite{Mamontov:06}). }
\label{fig6}
\end{figure}

We note that fast inelastic processes (such as phononsÉ) are far outside the dynamic range of the current QENS experiment. As a result, only the following elastic (all buried within the parameter $p_1$) and quasielastic scattering (with a relative spectral weight $(1-p_1)$) components contribute to the observed spectra in our hydrated sample:
{\begin{enumerate}
\item Scattering from the mobile molecules in the confined H$_2$O, quasielastic and some elastic (due to the confinement effect). Because the scattering cross-section of hydrogen is largely incoherent, the $Q$-dependence of the overall signal from the confined H$_2$O is entirely isotropic, with a relative elastic fraction (the true EISF) that varies with $Q$. 
\item Scattering from the SWNT matrix, all elastic. Because carbon scatters neutrons coherently, the $Q$-dependence of the signal from carbon has strong maxima at low $Q$ and at the position of the structural maximum ($\sim$ 0.4 {\AA}$-1$), which we intentionally avoided in our analysis.  The contributions from the dry sample can thus be conveniently accounted from model fits to the data, although not exclusively.
\item Possible scattering from water molecules in direct contact with the matrix walls that are immobile on the QENS time scale. This signal is elastic and roughly isotropic, again because of the dominant incoherent scattering by hydrogen. 
\end{enumerate}}

Regardless of the other contributions, it is only the scattering from the mobile molecules in the confined H$_2$O that yields QENS broadening, whereas other contributions are found in the elastic signal only (i.e. SWNT, immobile water molecules in contact with the pore walls, etcÉ).  Furthermore, the contribution (all elastic and coherent) from the carbon scattering to our data is further minimized due to the fact that the data exclude the structural maximum. While the elastic signal can be globally quantified with a single parameter $p_1$ without measuring the \lq dryÕ sample, knowledge of the relative spectral weight of the different components is however lost with this approach.  This analysis method is nevertheless well suited for studies of QENS linewidths. 

\section{Summary}
 We have investigated the effects of pressure on the dynamics of water molecules adsorbed in 16{ \AA} carbon nanotubes using neutron spectroscopy. The high resolution data reveal the presence of two diffusion processes, consistent with an inter- and intra-\lq water cage' dynamics. At full pore filling, the overall molecular dynamics is hindered by pressure. This effect is appreciably larger on the inter-\lq cage' dynamics than it is on the intra-\lq cage', but weaker than anticipated because the pressure inside the pores is anisotropic and probably affects just a small portion of molecules, seen by the neutrons.  
 
Recent molecular simulations \cite{Miyahara:00,Coasne:09} indicate that the pressure of water inside SWNT varies approximately exponentially with the bulk pressure. In this event, we anticipate that any fairly modest change in external pressure will significantly alter the molecular diffusion inside the pores, in qualitative agreement with our experiment.  Since the neutrons measure the global dynamics of the molecules adsorbed inside the pores, the net observed experimental effect appears to be less than it is at molecular level in some part of the sample.  Investigating the effect of pressure with other pore fillings, that could be used for example to separate the neutron response of water near the pore wall from that of water in the middle of the pore, or at higher pressure (5 or 10 kbar), is likely to provide valuable supplemental information, which would either confirm or refute the predictions. Grand Canonical Monte Carlo (GCMC) simulations with bond order analysis is being carried out \cite{Palmer:12,Long:11,Sliwinska:12} for water in SWNT carbons to predict the pressure tensor and phase transitions for confined water. These results are expected to provide a guide to experimental conditions where interesting phenomena are likely.

\section{Acknowledgement}
It is a pleasure to acknowledge S. Elorfi, R. Mills, and M. Loguillo at SNS for valuable technical support. We acknowledge stimulating discussion with A. Kolesnikov. Work at ORNL and SNS is sponsored by the Scientific User Facilities Division, Office of Basic Energy Sciences, US Department of Energy. JCP and KEG thank the NSF for support under grant CBET-1160151.

\section{Appendix}

\end{document}